\documentclass[fleqn,usenatbib,twocolumn]{mnras}

\usepackage{amssymb}
\usepackage{amsmath}
\usepackage{graphicx}
\usepackage{xcolor,ulem}
\usepackage{multirow}

\title[Effective coefficients for relativistic shocks]{Relating quasi-stationary one zone emission models to expanding relativistic shocks}

\author[E. Derishev]{Evgeny  Derishev \\
Institute of Applied Physics, 46 Ulyanov st., 603950 Nizhny Novgorod, Russia }

\date{Accepted XXX. Received YYY; in original form ZZZ}

\pubyear{2022}

\begin{document}

\newcommand{\diff}[1]{\mathop{}\!\mathrm{d} #1}

\label{firstpage}
\pagerange{\pageref{firstpage}--\pageref{lastpage}}
\maketitle

\begin{abstract}
For an expanding spherical relativistic shock, we derive relations between the parameters of downstream emitting zone and the quantities measured by a distant observer. These relations are formulated in terms of dimensionless effective coefficients combined with self-evident dimensional estimates. 
Our calculations take into account evolution of the shock's Lorentz factor, geometrical delay due to the shock's front curvature, and angular dependence of Lorentz boost for frequency and brightness.
The relations are designed primarily for application in Gamma-Ray Burst afterglow studies, although they may have a broader use.
\end{abstract}


\section{Introduction} 
\label{sec:intro}

Emission observed at the afterglow stage of Gamma-Ray Bursts (GRBs) comes from the downstream region of expanding relativistic shock (see, e.g. \cite{Afteglow1,Afteglow2} for introduction into the afterglow concept and \cite{GRBreview} for a GRB review). 
This kind of geometrical setup --- emitting region that expands at nearly the speed of light --- poses additional challenge for theory. 
Even if one understands local conditions in the emitting region and is able to calculate the local radiation spectrum, one also needs to know how these quantities transform into quantities measured by a distant observer. 
This is a non-trivial problem and the difficulty is exactly in relativistic motion towards the observer, so that time delays for shock's expansion are comparable to time delays for light travel. 

Because of Lorentz boost, the radiation produced in the emitting region is strongly beamed in the direction of shock's motion, and a distant observer effectively sees only a small segment, that subtends angle of the order of $1/\Gamma_\mathrm{d}$, where $\Gamma_\mathrm{d}$ is the Lorentz factor of the shocked gas. 
Even then, different parts of this small segment, being observed simultaneously, significantly differ in ages and hence in distance from the origin. 
As a result, a spherical shock appears in strongly elongated shape and the regions that are further away from the line of sight are closer to the origin and younger.
These geometrical effects and some of their consequences are discussed in \cite{Waxman97}, \cite{PanaitescuMeszaros98}, and \cite{AfterglowRing}.

A model of GRB afterglow emission has to take into account all geometric, delay, and Lorentz boost effects due to expanding relativistic shock. Formulating the problem in minimalistic (one-zone) and most abstract way, we need to relate the following quantities. 
(1) Effective radius of the emitting zone to observer's time.  
(2) Particles' lifetime in the emitting zone to observer's time.  
(3) Photon's energy in the emitting zone to observed photon's energy.
(4) Radiation energy density in the emitting zone to isotropic equivalent luminosity of the source.
For all of these relations there are fairly obvious scaling laws (dimensional relations), that are to be complemented by dimensionless coefficients. Different authors estimate these coefficients in different ways that result in different numbers (see references in Table~\ref{tab:coefficients}). 
Moreover, the only paper that gives, either explicitly or implicitly, the complete set of coefficients is \cite{Afteglow2}, where estimates are limited to the case of adiabatic shock propagating into constant-density external medium. The goal of this paper is to calculate consistently all the coefficients for both constant-density and wind-like external medium, as well as generalize the results for the case of partially radiative shocks.

It should be noted that a particular choice of the effective coefficients does not affect ability to model GRB afterglow emission: as demonstrated in \cite{SEDmodel}, different sets of the coefficients result in equally good spectral fits, but they correspond to different parameters of the emitting zone (the magnetic field strength, Lorentz factor of injected electrons and their energy fraction, etc). 
The reverse is not true. Different sets of the coefficients produce dissimilar spectra with the same emission zone parameters. It is therefore a daunting, nearly impossible, task to compare  numerical models of GRB afterglow emission if the authors do not specify their choice of effective coefficients. So, having a commonly accepted convention which coefficients to pick in which situation is critical in order to be able to cross-check various numerical spectral fits.

To calculate the aforementioned coefficients, we make three model assumptions. 
(1) The shock is spherical --- this is a good approximation for jetted explosions as long as the jet's opening angle is larger than $1/\Gamma_\mathrm{d}$. 
(2) The emission comes from a geometrically thin layer at the shock's front --- this approximation is good in the fast cooling regime (see also discussion of this assumption for the slow cooling regime in Sect.~\ref{sec:discuss}). 
(3) The fraction of shock's power that is converted to radiation is constant. 
The only theoretical model that gives specific predictions about assumptions (2) and (3) is the pair-balance model \citep{PairBalance}, and within its framework these assumptions hold. 

Though our analysis was made with application to GRB afterglows in mind, we note that a similar geometry of relativistically expanding emitting zone may form as a result of explosive power increase in various other sources that harbor relativistic outflows, such as Active Galactic Nuclei, microquasars and pulsars.

\section{Formulation of the problem} 
\label{sec:formulation}

Our analysis is based on the self-similar hydrodynamic solution for decelerating relativistic blast wave obtained by \cite{BlandfordMcKee}. We follow their notation for the external density profile, $\rho \propto R^{-k}$, and for the shock deceleration law, $\Gamma_\mathrm{fr}^2 \propto R^{-m}$, where $\Gamma_\mathrm{fr} (R)$ is the Lorentz factor of the shock front as a function of distance from the center of explosion. 
An adiabatic blast wave has $m=3$ for constant-density medium and $m=1$ for wind density profile. Partially radiative shocks decelerate faster, but also obey  power-law if the fraction of radiated energy, $\epsilon_\mathrm{r}$, is constant \citep{PartiallyRadiativeShock}.
Although the self-similar solution describes spherical shock, it remains approximately valid if the shock subtends an angle $\gg 1/\Gamma_\mathrm{fr}$. 

It is important to distinguish between the shock's front Lorentz factor and the Lorentz factor of shocked material immediately downstream of the shock. 
The former characterizes shock's motion whereas the latter characterizes the Doppler boost for radiation produced in the emitting zone.
Jump conditions for adiabatic relativistic shock give $\Gamma_\mathrm{d} = \Gamma_\mathrm{fr}/\sqrt{2}$. This relation needs a correction when the radiative efficiency $\epsilon_\mathrm{r}$ is large.

The radius of the most faraway part of the shock as seen by a distant observer at time $t_\mathrm{obs}$ is $R_\mathrm{sh}$. This part of the shock propagates along the line of sight, so its signal arrives first. Signals from other parts of spherical shock, that move at some angles to the line of sight, are delayed by the time it takes light to propagate from more distant segments of the sphere. For the observer, these segments appear at smaller separation from the explosion's center as if they had less time to propagate (see depiction of the equal arrival time surface in Fig.~\ref{EAT_surface_cartoon}).
We will use subscript ``sh'' to denote quantities that refer to the head-on part of the shock at distance $R_\mathrm{sh}$, such as $\Gamma_\mathrm{sh} \equiv \Gamma_\mathrm{fr} (R_\mathrm{sh})$, $\rho_\mathrm{sh} \equiv \rho (R_\mathrm{sh})$, etc.

Aiming at description of emission from a relativistic spherical shock, we need to establish relations between model-independent quantities (observer's time since explosion $t_\mathrm{obs}$ and the Lorentz factor of head-on portion of the shock front at this time $\Gamma_\mathrm{sh}$) and the parameters of one-zone radiation model:
\begin{subequations}
\label{CoefficientDefinitions}
\begin{align} 
&    R_\mathrm{em} = C_\mathrm{_R} \Gamma_\mathrm{sh}^2 c {t_\mathrm{obs} / (1+z)}  \label{emission_radius} \\
&    t_\mathrm{eff} = {C_\mathrm{t} \Gamma_\mathrm{sh} {t_\mathrm{obs} / (1+z)}} \label{effective_time} \\   
&    h\nu_\mathrm{obs}  = C_{_\Gamma} \Gamma_\mathrm{sh} h\nu {/ (1+z)} \\
&    L = C_\mathrm{_L} \epsilon_r \; (1+z) E_\mathrm{kin}/t_\mathrm{obs}  \label{luminosity} \\
&    E_\mathrm{kin} = C_\mathrm{_E} \Gamma_\mathrm{sh}^2 M c^2 \ , 
\end{align}
\end{subequations}
Here $t_\mathrm{eff}$ is the effective particles' lifetime in the emitting zone, $R_\mathrm{em}$ its effective radius (which is smaller than $R_\mathrm{sh}$) and $h\nu_\mathrm{obs}$ the energy of observed photons, Lorentz boosted from $h\nu$ in the emitting zone comoving frame, $M$ the swept-up mass, $L$ the shock's bolometric luminosity, and $\epsilon_\mathrm{r}$ the fraction of shocked matter energy that is dissipated to radiation (we will call this fraction radiative efficiency).
In the fast cooling regime it approximately equals the fraction of energy that goes into accelerated particles, $\epsilon_\mathrm{r} \approx \epsilon_\mathrm{e}$, whereas in the slow cooling regime $\epsilon_\mathrm{r} \ll \epsilon_\mathrm{e}$.

The dimensionless factors $C_\alpha$ in these scaling relations are not immediately obvious if one wants to incorporate averaging along equal arrival time surfaces.  
Below we calculate their values and give rationale for our choice of the averaging procedure.

When evaluating the coefficients $C_\alpha$ we can assume zero redshift ($z=0$) without loss of generality. For this reason, throughout the rest of this paper the words ``distant observer'' refer to an observer located at the same redshift as the explosion's progenitor.

For completeness we also introduce the image width factor that relates the transverse size of the shock in the image plane to the observer's time. 
For a distant observer at the same redshift
\begin{equation} \label{width}
    w = C_\mathrm{w} \Gamma_\mathrm{sh} c t_\mathrm{obs} \ .
\end{equation}
This relation is not used in models of the emission zone, but may be useful for interpreting observations.

The rest of the paper is organized as follows. 
We discuss in Sect.~(\ref{sec:EATsurface}) the shape of the shock as it appears for a distant observer taking into account light delay and aberration effects, and calculate $C_\mathrm{w}$. 
Individual coefficients are calculated in the following sections: $C_\mathrm{_L}$ in Sect.~(\ref{sec:luminosity}),
$C_\mathrm{_R}$ in Sect.~(\ref{sec:size}), $C_{_\Gamma}$ in Sect.~(\ref{sec:boost}), and $C_\mathrm{t}$ in Sect.~(\ref{sec:time}).
A reader who is not interested in the technical details may skip directly to Sect.~(\ref{sec:discuss}), where we summarize (in Table~\ref{tab:coefficients}) the numerical values of the coefficients that were introduced in Sect.~(\ref{sec:formulation}) and discuss the results.

\section{Surface of equal arrival time}
\label{sec:EATsurface}

\begin{figure}
	\includegraphics[width=1.0\columnwidth]{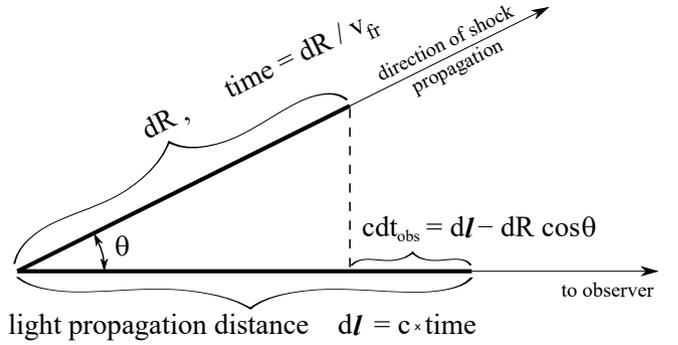}
    \caption{A cartoon explaining the relation between distance $\diff{R}$, travelled by the shock's front, and observer's time change $\diff{t_\mathrm{obs}}$ (see Eq.~\ref{dR(dt)}).}
    \label{dR}
\end{figure}

\begin{figure}
	\includegraphics[width=1.0\columnwidth]{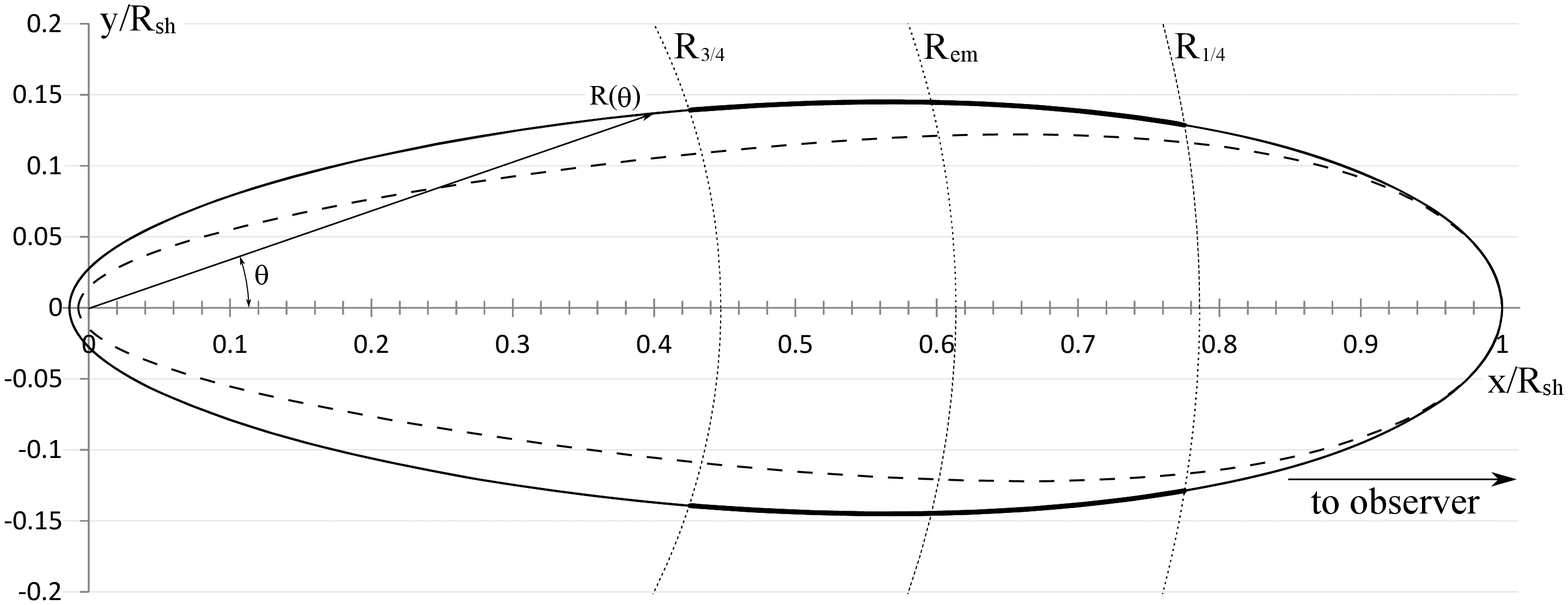}
    \caption{{\it Solid line:} shape of the surface of equal arrival time calculated for $\Gamma_\mathrm{sh}=3$ and $m=1$ (adiabatic shock in wind-like density profile). The distances marked as $R_{1\!/\!4}$, $R_\mathrm{em}$, and $R_{3\!/\!4}$ delimit regions that contribute $1/4$, $1/2$, and $3/4$ to the total bolometric luminosity, counting from the most faraway point to the origin (see Sec.~\ref{sec:size}). The part between $R_{1\!/\!4}$ and $R_{3\!/\!4}$, shown in thick line, is responsible for a half of the shock's bolometric luminosity.
    {\it Dashed line:} the equal arrival time surface calculated for comparison with the same $\Gamma_\mathrm{sh}=3$, but with $m=3$ (adiabatic shock in constant density profile). Note that $R_{1\!/\!4}$, $R_\mathrm{em}$, and $R_{3\!/\!4}$ are different in this case. Note also that $R_\mathrm{sh}$ for the dashed curve is two times larger and if both curves were plotted to the same scale, then they are tangent to each other at the point closest to the origin.}
    \label{EAT_surface_cartoon}
\end{figure}

For a shock moving at an angle $\theta$ to the line of sight, increase in the shock's front distance from the origin $\diff{R}$ is related to observer's time change as (see Fig.~\ref{dR})
\begin{equation} \label{dR(dt)}
    \diff{R} = \frac{\beta_\mathrm{fr} c}{1-\beta_\mathrm{fr} \cos\theta} \diff{t_\mathrm{obs}} \, .
\end{equation}
In the relativistic limit (i.e. with $O\left(\Gamma^{-2} \right)$ precision)
\begin{equation} \label{cos_in_rel_limit}
    1-\beta \cos\theta \equiv \left(1-\beta \right) + \beta \left(1-\cos\theta \right)
    \simeq \frac{1}{2 \Gamma^2} + \left(1-\cos\theta \right) 
\end{equation}
and, integrating $\diff{t_\mathrm{obs}}$ from Eq.~(\ref{dR(dt)}) with substitution $\Gamma_\mathrm{fr} \left( R^\prime \right) =  \left( R_\mathrm{sh} /R^\prime \right)^{m/2}  \Gamma_\mathrm{sh}$, we find that 
\begin{multline} \label{EAT_equation}
    t_\mathrm{obs} = \frac{1}{c} \int_0^{R} \left( \frac{\left( R^\prime \right)^{m}}{2 \left(R_\mathrm{sh} \right)^{m} \Gamma_\mathrm{sh}^2} + 1-\cos\theta \right)  \diff{R^\prime} \\
    =  \frac{1}{c} \left( \frac{R^{m+1}}{2 \left( m+1 \right) \left(R_\mathrm{sh} \right)^{m} \Gamma_\mathrm{sh}^2} + \left( 1-\cos\theta \right) R \right)  
    \ .
\end{multline}
This equation defines the surface such that photons emitted along it reach the distant observer at the same time. The most distant from the center of explosion point along the equal arrival time surface is at zero observation angle, where
\begin{equation} \label{Rsh}
    R_\mathrm{sh} \equiv R(\theta=0) = 2 \left( m+1 \right)  \Gamma_\mathrm{sh}^2 c t_\mathrm{obs} \ .
\end{equation}
For further use, it is convenient to re-write Eq.~(\ref{EAT_equation}) in terms of dimensionless distance $r \equiv R/R_\mathrm{sh}$:
\begin{equation} \label{dimensionless_EAT_equation}
    r^{m+1} + 2 \left( m+1 \right) \Gamma_\mathrm{sh}^2 \left( 1-\cos\theta \right) r = 1    \ .
\end{equation}
Equation (\ref{dimensionless_EAT_equation}) readily solves with respect to $\theta$ and in some special cases (most notably in the case of $m=1$, i.e. adiabatic shock propagating into wind-like density profile) it also solves analytically with respect to $r$. The shape of equal arrival time surface is shown in Fig.~(\ref{EAT_surface_cartoon}), where low value $\Gamma_\mathrm{sh}=3$ is chosen for illustrative purposes. With typical for GRB afterglows $\Gamma_\mathrm{sh} \sim$~tens, the equal arrival time surface is more like a needle with its blunt end pointing towards the observer.

As seen from Fig.~(\ref{EAT_surface_cartoon}), the brightest part of the equal arrival time surface appears to the distant observer as a hollow cylinder. Therefore, in projection to the image plane the shock looks like a narrow bright ring with much dimmer interior. 
The apparent radius of shock's image is given by the maximum value of $y \equiv r \sin\theta$ from Eq.~(\ref{dimensionless_EAT_equation}). In the limit $\Gamma_\mathrm{sh} \gg 1$, when the relevant angles are small ($\theta \ll 1$), the equation simplifies to
\begin{equation} \label{approximate_EAT_equation}
     \left( m+1 \right) \Gamma_\mathrm{sh}^2 \ y^2 = r - r^{m+2}   \ ,
\end{equation}
so that $r \left( y_\mathrm{max} \right) = \left( m+2 \right)^{-1/(m+1)}$. Substituting  $r \left( y_\mathrm{max} \right)$ into Eq.~(\ref{approximate_EAT_equation}) gives
\begin{equation} \label{C_w}
       C_\mathrm{w} = 2 \left( m+1 \right) \left( m+2 \right)^{-\frac{m+2}{2(m+1)}}   \ .
\end{equation}
This result holds for both adiabatic and partially radiative blast waves. Numerical values of $C_\mathrm{w}$ for some cases are given in Table~(\ref{tab:coefficients}).

\section{Luminosity coefficient}
\label{sec:luminosity}

The bolometric luminosity of an expanding shock is the sum of contributions from shock's elements located along the equal arrival time surface. Our approach to calculating the luminosity coefficient $C_\mathrm{L}$ is straightforward: find apparent luminosity for shock's elements, integrate over the equal arrival time surface, and express the result in terms of dimensional quantities as in Eq.~\ref{luminosity}. Two assumptions are made on the way: (1) the radiation is isotropic in the fluid comoving frame and (2) all radiation comes from thin region in the downstream immediately adjacent to the shock, thus ignoring extent of the emitting zone into the downstream. Both assumptions stem from one-zone formulation of the problem that postulates uniform emitting zone, where there is no reason for anisotropy to emerge and there is no way to distinguish between particles with different radiative cooling time (see also  discussion in Sect.~\ref{sec:discuss}). 

Consider a shock's front element subtending solid angle $\diff{\Omega}$. It sweeps a mass $\diff{M} = \rho R^2 \diff{\Omega} \diff{R}$ while advancing a distance $\diff{R}$. 
The energy associated with swept mass is $\Gamma_\mathrm{d} \diff{M} c^2$ in the fluid comoving frame, and a fraction $\epsilon_\mathrm{r}$ of this energy is converted into radiation. For a shock moving at an angle $\theta$ to the line of sight, the distance $\diff{R}$ is related to observer's time change through Eq.~(\ref{dR(dt)}) (see also Fig.~\ref{dR})
so that the energy release rate (in the fluid comoving frame) is 
\begin{equation} \label{energy_release_rate}
    \lambda(R,\theta) 
    = \frac{\epsilon_\mathrm{r} \Gamma_\mathrm{d} \diff{M} c^2}{\diff{\Omega} \diff{t_\mathrm{obs}}}
    = \epsilon_\mathrm{r} \rho c^3 R^2 \frac{\Gamma_\mathrm{d} \beta_\mathrm{fr}}{1-\beta_\mathrm{fr} \cos\theta} \ . 
\end{equation}

The apparent luminosity of a moving source is Doppler boosted with respect to the luminosity in the proper frame,
\begin{equation} \label{luminosity_boost}
    L_\mathrm{app} = \delta^3 L_\mathrm{prop} \ ,
\end{equation}
where
\begin{equation} \label{Doppler}
    \delta = \frac{1}{\Gamma_\mathrm{d} \left(1-\beta_\mathrm{d} \cos\theta \right)}
\end{equation}
is the Doppler factor for photons observed at the angle $\theta$ to the shock's normal. Note that the relation~(\ref{luminosity_boost}) is different from the usual proportion $L_\mathrm{app} \propto \delta^4$ because the velocity of emitting material $\beta_\mathrm{d}$, that is responsible for photon beaming and blueshift, differs from the velocity $\beta_\mathrm{fr}$, at which the source propagates. 
In Eq.~(\ref{luminosity_boost}), two powers of $\delta$ are from photon beaming and one power is from photons' energy boost. 
The time contraction multiplier is taken into account separately in Eq.~(\ref{dR(dt)}), where $\diff{R}/\diff{t_\mathrm{obs}}$ is the largest when the shock front moves at zero angle to the line of sight.

The bolometric luminosity of the shock, integrated over the whole emitting zone, is 
\begin{equation} \label{full_luminosity}
    L (t_\mathrm{obs}) = \oint \lambda \delta^3 \diff{\Omega} 
    = 2\pi \int_{-1}^{1} \lambda \delta^3 \diff{\cos\theta}     \, .
\end{equation}
Using relativistic approximation for $1-\beta \cos\theta$ from Eq.~(\ref{cos_in_rel_limit}) and then substituting $\left( 1 - \cos \theta \right)$ from Eq.~(\ref{dimensionless_EAT_equation}), we find that
\begin{equation} 
    \frac{1}{1-\beta \cos\theta} 
    \simeq \frac{2 (m+1) \Gamma^2 r}{(m+1) r + \left( \Gamma^2 / \Gamma_\mathrm{sh}^2 \right)  \left( 1 - r^{m+1} \right)} \, .
\end{equation}
Finally, taking into account that $\rho = r^{-k} \rho_\mathrm{sh}$, $\Gamma_\mathrm{d} = \Gamma_\mathrm{fr} / \sqrt{2}$ (requires adiabatic jump conditions at the shock front, i.e. $\epsilon_\mathrm{r}$ is not large), and $\Gamma_\mathrm{fr}^2 = r^{-m} \Gamma_\mathrm{sh}^2$ (requires that $\epsilon_\mathrm{r}$ is either negligibly small or constant), we obtain
\begin{equation} \label{lambda_term}
    \lambda \simeq  
    \sqrt{2} (m+1) \epsilon_\mathrm{r} \rho_\mathrm{sh} c^3 R_\mathrm{sh}^2\Gamma_\mathrm{sh}^3 \frac{r^{2-k-3m/2}}{m + r^{-(m+1)} }  
\end{equation}
and
\begin{equation} \label{delta_term}
    \delta \simeq  
    2 \sqrt{2} (m+1) \Gamma_\mathrm{sh} \frac{r^{-m/2}}{2m + 1 +  r^{-(m+1)}} \ .
\end{equation}

Now, the integrand in Eq.~(\ref{full_luminosity}) can be written as function of $r$, and we need to express the differential $\diff{\cos\theta}$ through $\diff{r}$. This can be done by implicitly differentiating Eq.~(\ref{dimensionless_EAT_equation}) and then substituting $\left( 1 - \cos \theta \right)$ from the same equation:
\begin{multline} \label{differential_term}
    \left[ (m+1) r^{m} + 2 (m+1) \Gamma_\mathrm{sh}^2 \left( 1 - \cos \theta \right) \right] \diff{r} 
    = 2 (m+1) \Gamma_\mathrm{sh}^2 r \diff{\cos\theta} \\
    \quad \Rightarrow \quad
    \diff{\cos\theta} =  \frac{1 + m r^{m+1}}{2 (m+1) \Gamma_\mathrm{sh}^2 r^2} \diff{r} \, . 
\end{multline}

\begin{figure}
	\includegraphics[width=1.0\columnwidth]{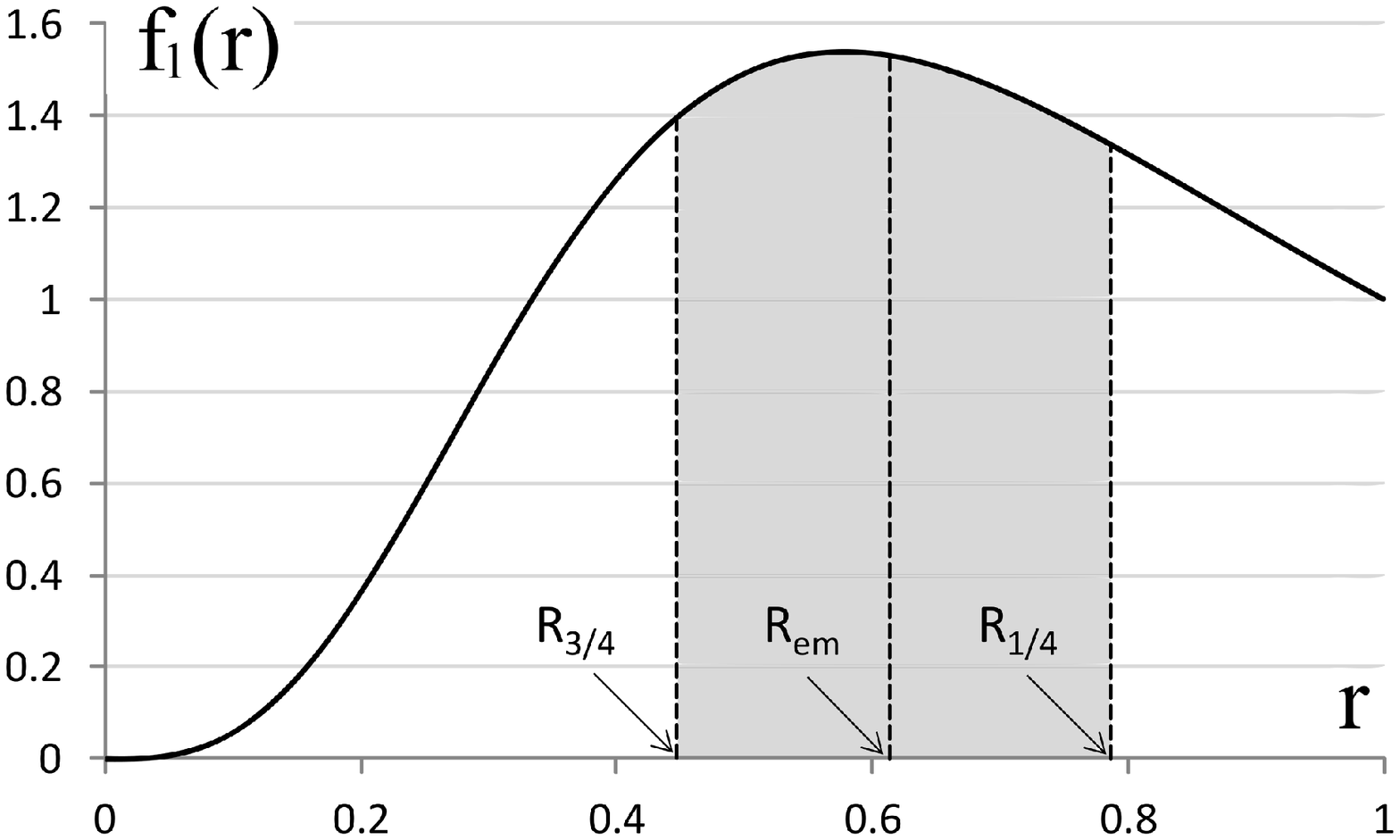} \\
	\includegraphics[width=1.0\columnwidth]{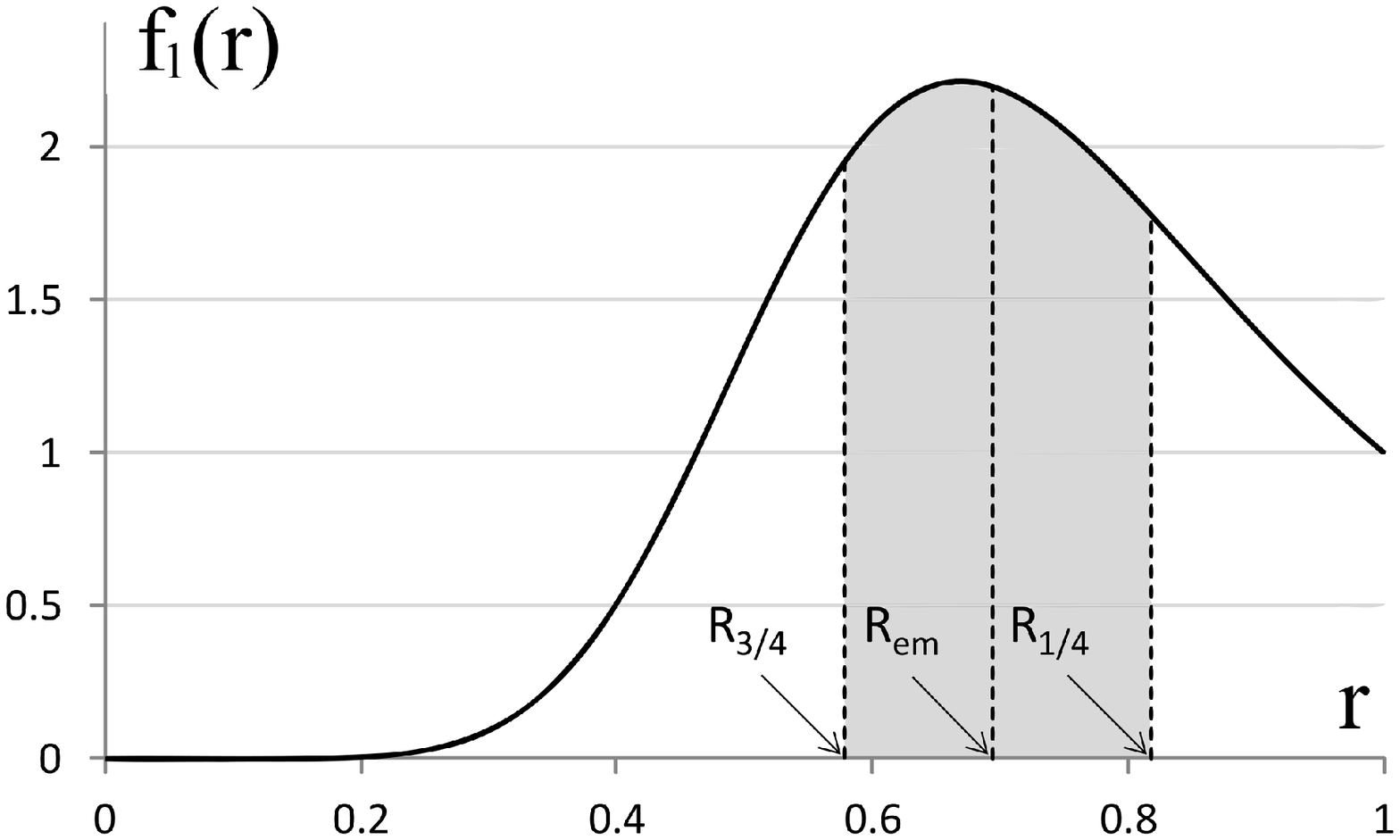}
    \caption{Contributions to the shock's bolometric luminosity from different radii (Eq.~\ref{integrand_f1}). The points marked as $R_{1\!/\!4}$, $R_\mathrm{em}$, and $R_{3\!/\!4}$ delimit regions that contribute $1/4$, $1/2$, and $3/4$ to the total bolometric luminosity, counting from the most faraway point to the origin. The shaded region gives one half of the luminosity.
    {\it Top panel:} adiabatic shock, propagating into wind-like density profile ($m=1$, $k=2$).
    {\it Bottom panel:} adiabatic shock, propagating into constant density profile ($m=3$, $k=0$).
    }
    \label{LuminosityDistribution}
\end{figure}

Collecting the terms from Eqs.~(\ref{lambda_term},\ref{delta_term},\ref{differential_term}) we rewrite Eq.~(\ref{full_luminosity}) as 
\begin{equation} \label{luminosity(r)_simple}
    L (t_\mathrm{obs}) \simeq 4\pi \epsilon_\mathrm{r} \rho_\mathrm{sh} c^3 R_\mathrm{sh}^2 \Gamma_\mathrm{sh}^4 
   \,I_1\!\!\left( m,k \right)     \ ,
\end{equation}
where
\begin{equation} \label{luminosity(r)_integral_}
   I_1\!\!\left( m,k \right)  =
    \int_0^1   f_1\!(r) \diff{r}   
\end{equation}
and
\begin{equation} \label{integrand_f1}
    f_1\!(r) = 8 \left( m+1 \right)^3  r^{1-k-2m} \left( 2m + 1 +  r^{-(m+1)} \right)^{-3}      \,  . 
\end{equation}
When intergating over $\diff{r}$, we replace the lower limit $r_\mathrm{min} \simeq 1/\left( 4 \left( m+1 \right) \Gamma_\mathrm{sh}^2 \right)$ by 0. The integrand $f_1\!(r)$ is shown in Fig.~(\ref{LuminosityDistribution}).

In the case of partially radiative shock, where $m > 3-k$, the integral in Eq.~(\ref{luminosity(r)_integral_}) is to be evaluated numerically.
In the case of adiabatic shock, where $m=3-k$, integration (by substitution $\xi \equiv r^{-(m+1)}$) is trivial, $I_1\!\!\left( m,k \right)  = 1$, and Eq.~(\ref{luminosity(r)_simple}) becomes
\begin{equation} \label{luminosity_adiabatic}
    L (t_\mathrm{obs}) \simeq 4\pi \epsilon_\mathrm{r} \rho_\mathrm{sh} c^3 R_\mathrm{sh}^2 \Gamma_\mathrm{sh}^4 
   \quad \left( \mathrm{for} \; m = 3-k \right)  .
\end{equation}
Finally, expressing the swept-up mass as $M = 4 \pi \rho_\mathrm{sh} R_\mathrm{sh}^3 / \left( 3-k \right)$ and grouping terms in Eq.~(\ref{luminosity(r)_simple}) appropriately, we arrive at
\begin{equation} \label{C_L}
    C_\mathrm{_L}    =      \frac{\left( 3-k \right)  \,I_1\!\!\left( m,k \right)}{2 \left( m+1 \right)   C_\mathrm{_E}}  \ .
\end{equation}

Values of the luminosity coefficients $C_\mathrm{_L}$ calculated for some special cases are summarized in Table~(\ref{tab:coefficients}).

\begin{figure}
	\includegraphics[width=1.0\columnwidth]{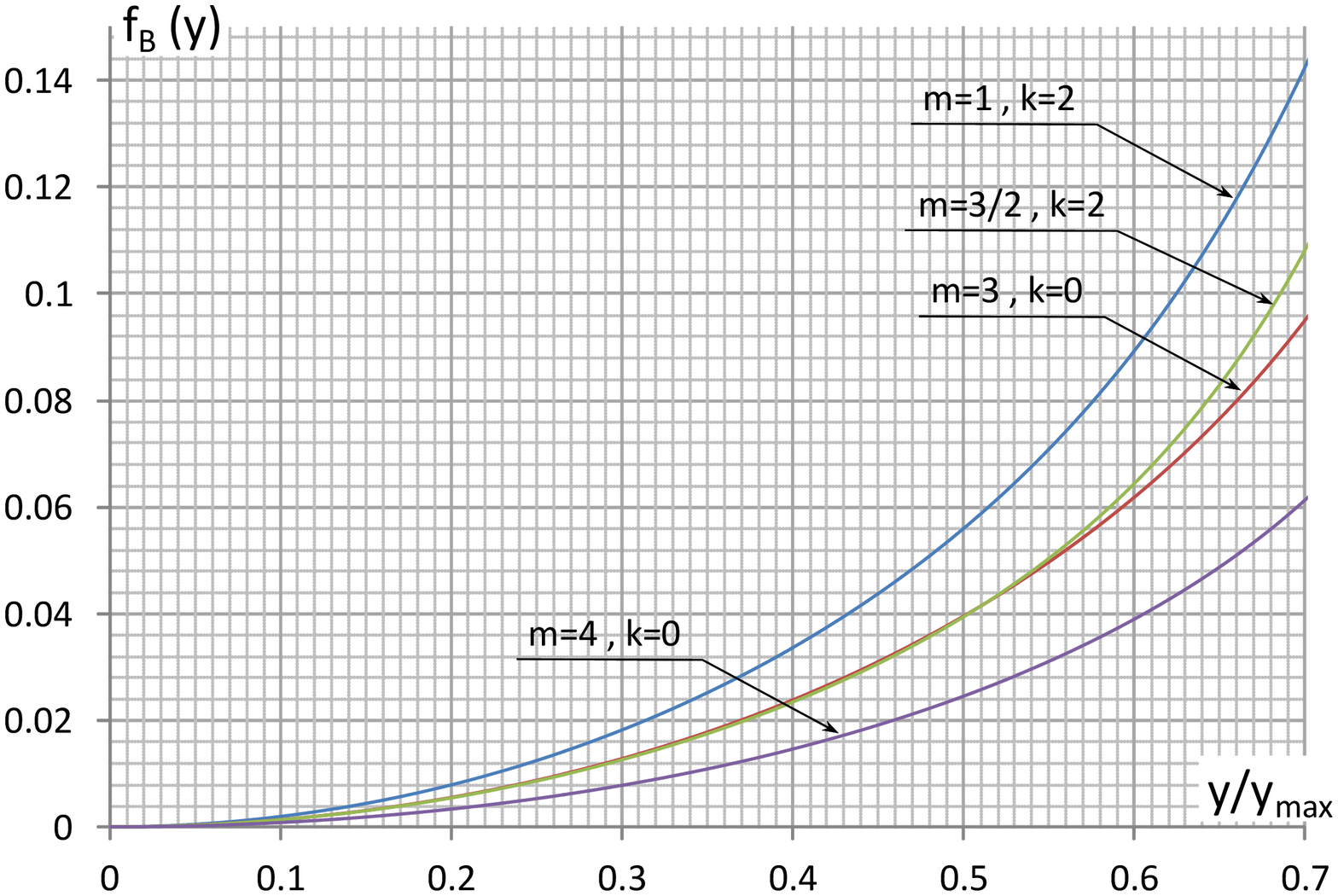} \\
	\includegraphics[width=1.0\columnwidth]{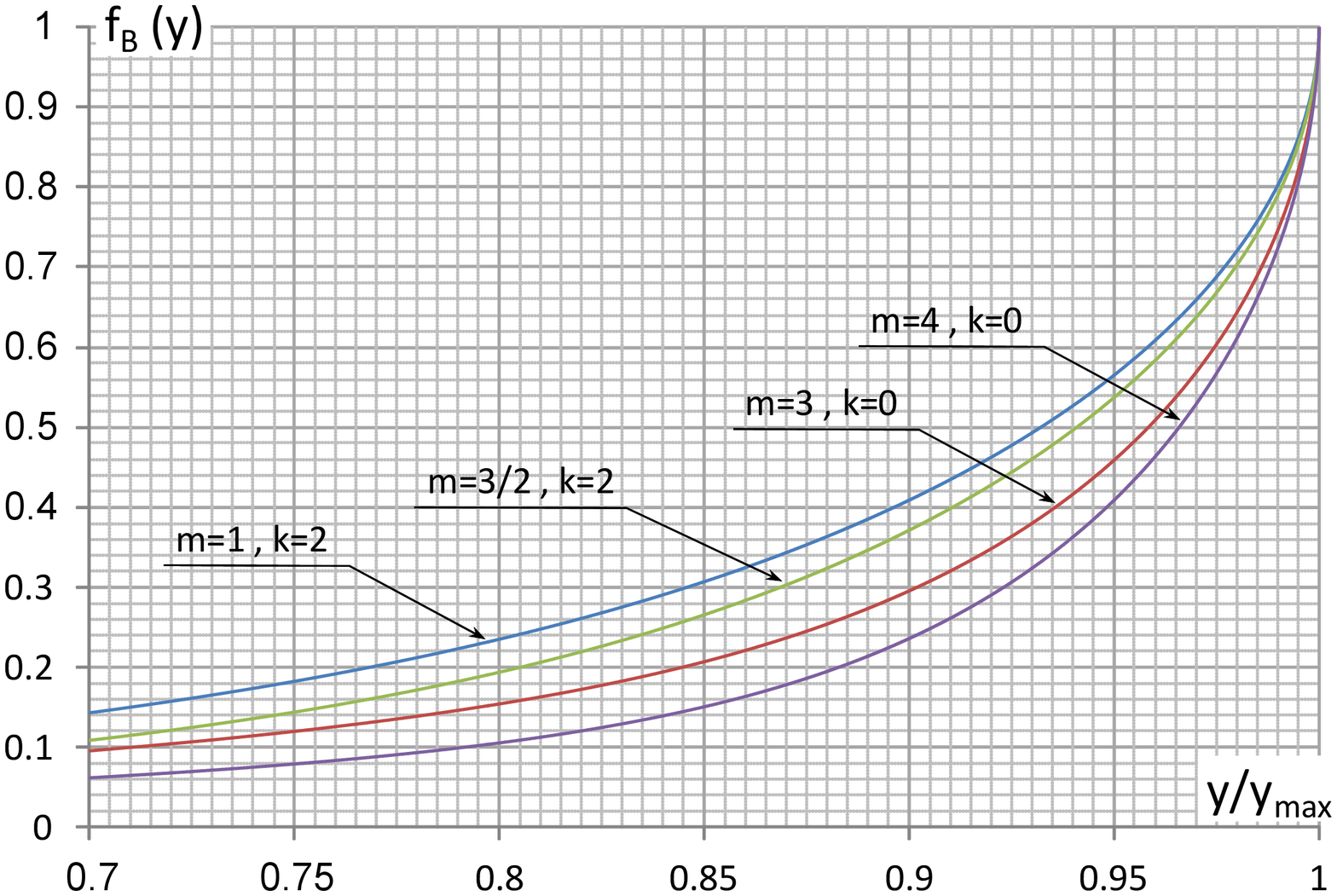}
    \caption{Fraction of bolometric luminosity coming from the central part of shock's image inside radius $y$ as function of this radius. The four curves are for shocks propagating into wind-like density profile, both adiabatic ($m=1,k=2$) and partially radiative ($m=3/2,k=2$), and into constant-density medium, both adiabatic ($m=3,k=0$) and partially radiative ($m=4,k=0$).
    Note that the values of $y_\mathrm{max}$ are different for all curves and so are their actual spacial scales.
    {\it Top panel:} the leftmost part of the plot, which covers the inner half of the shock's image.
    {\it Bottom panel:} the rightmost part of the plot, which covers the outer half of the shock's image.
    }
    \label{fig:f_B}
\end{figure}

The distribution of shock's brightness in projection to the image plane is shown in Fig.~(\ref{fig:f_B}), where we plot the fraction of bolometric luminosity from a disk inside radius $y$ as function of this radius,
\begin{equation} \label{f_B}
    f_\mathrm{_B} (y)   = \left. \left( \int_0^{r_1} f_1\!(r) \diff{r} + \int_{r_2}^1 f_1\!(r) \diff{r} \right) \middle/ I_1\!\!\left( m,k \right) \right. \ ,
\end{equation}
where $r_1 (y) < r_2 (y)$ are the two roots of Eq.~(\ref{approximate_EAT_equation}). The bolometric luminosity is dominated by a narrow (just few percent wide) ring at the image's edge. The faster shock decelerates the more contracted is this ring. This strongly pronounced limb brightening is partially due to projection effect itself and partially due to the peak of radial brightness distribution (Eq.~\ref{integrand_f1}) being adjacent to the image's rim (they exactly coincide for adiabatic shocks).

\section{Size coefficient} 
\label{sec:size}

We adopt such a value for $C_{_\mathrm{R}}$ that corresponds to the median emission radius, i.e. at a given observer's time half of the luminosity is due to parts of the blast wave observed at distances smaller than $R_\mathrm{em}$, and the parts observed at distances larger than $R_\mathrm{em}$ sum up to give the other half.
In terms of dimensionless radius, $r_\mathrm{em}$ satisfies the following equation
\begin{equation}  \label{R_em}
    \int_0^{r_\mathrm{em}}  f_1\!(r)  \diff{r}     = \frac{1}{2}   I_1\!\!\left( m,k \right)  \ . 
\end{equation}
For an adiabatic blast wave (i.e. $m=3-k$), the above equation evaluates to 
\begin{multline} \label{R_em-adiabatic}
    \left( 2 m +1  + r_\mathrm{em}^{-(m+1)} \right)^{-2}  = \frac{1}{8(m+1)^2} \\
    \qquad \Rightarrow \qquad
    r_\mathrm{em} = \left( 2\sqrt{2} \left( m+1 \right) - 2m -1 \right)^{-\frac{1}{m+1}}    \, ,
\end{multline}
so that
\begin{equation}
    C_{_\mathrm{R}} = 2(m+1) \left( 2\sqrt{2} (m+1) - 2m -1 \right)^{-\frac{1}{m+1}}    \, . 
\end{equation}
For a partially radiative blast wave (i.e. $m>3-k$) one needs to integrate Eq.~(\ref{R_em}) numerically to find $R_\mathrm{em}$ and then $C_{_\mathrm{R}}$.
Values of the size coefficient $C_{_\mathrm{R}}$ calculated for some special cases are summarized in Table~(\ref{tab:coefficients}). 

In addition to $R_\mathrm{em}$, it is convenient to introduce two auxiliary distance marks, $R_\mathrm{1\!/\!4}$ and $R_\mathrm{3\!/\!4}$, that delimit regions contributing $1/4$ and $3/4$ of the total bolometric luminosity, counting from the most faraway point of the equal arrival time surface, i.e. 
\begin{equation} \label{R_1/4-R_3/4-definitions}
    \int_{r_\mathrm{1\!/\!4}}^1  f_1\!(r)  \diff{r}     = \frac{1}{4}   I_1\!\!\left( m,k \right)  \ , \quad 
    \int_{r_\mathrm{3\!/\!4}}^1  f_1\!(r)  \diff{r}     = \frac{3}{4}   I_1\!\!\left( m,k \right)  \ . 
\end{equation}

In the case of adiabatic shock, there are analytic expressions for these distances
\begin{equation} \label{R_1/4}
    r_\mathrm{1\!/\!4} = \left( 4 \left( m+1 \right) / \sqrt{3} - 2m -1 \right)^{-\frac{1}{m+1}}    \, .
\end{equation}
and
\begin{equation} \label{R_3/4}
    r_\mathrm{3\!/\!4}    = \left( 2m + 3 \right)^{-\frac{1}{m+1}}    \, .
\end{equation}
The values of $R_\mathrm{em}$, $R_\mathrm{1\!/\!4}$, and $R_\mathrm{3\!/\!4}$ are shown in Fig.~(\ref{LuminosityDistribution}) for adiabatic shocks. Normalized values of $r_\mathrm{1\!/\!4}$, and $r_\mathrm{3\!/\!4}$ for both adiabatic and partially radiative shocks are listed in Table~(\ref{tab:C_R_comparison}).

One could devise alternative definitions of the effective emission radius and ascribe it either to the point where the brightness distribution (\ref{integrand_f1}) takes its maximum value,
\begin{equation} \label{C_R-alt1}
r_\mathrm{p} = \left( \frac{m-k+4}{(2m+1)(2m+k-1)} \right)^{\frac{1}{m+1}}
\quad \Rightarrow \quad 
\widetilde{C}_\mathrm{_R} = 2(m+1) r_\mathrm{p} \ ,
\end{equation}
or to the brightness-weighthed average of $r$,
\begin{equation} \label{C_R-alt2}
\bar{r} = \left. \int_0^1 r f_1\!(r) \diff{r} \middle/ I_1\!\!\left( m,k \right) \right.
\quad \Rightarrow \quad 
\overline{C}_\mathrm{_R}  = 2(m+1) \bar{r} \ .
\end{equation}
The values that follow from alternative definitions of $C_\mathrm{_R}$ would place the effective emission radius at nearly the same location --- see Table~(\ref{tab:C_R_comparison}) for comparison.

\begin{table} 
\caption{The values of the size coefficient, that correspond to alternative definitions,
and normalized (to $R_\mathrm{sh}$) distances, that correspond to $1/4$ and $3/4$ shares of the bolometric luminosity.}
\label{tab:C_R_comparison}
\begin{tabular}{ccccccc}
\hline
    $k$ & $m$ & $C_\mathrm{_R}$ & $\widetilde{C}_\mathrm{_R}$ & $\overline{C}_\mathrm{_R}$ 
    & $r_{1\!/\!4}$ & $r_{3\!/\!4}$ \\
    \hline
    0 & 3 & $\approx 5.55$ & $\approx 5.35$ & $\approx 5.56$ 
    & $\approx 0.818$ & $\approx 0.577$     \\
    \hline  
    0 & 4 & $\approx 6.87$ & $\approx 6.62$ & $\approx 6.92$ 
    & $\approx 0.797$ & $\approx 0.587$     \\
    \hline  
    2 & 1 & $\approx 2.45$ & $\approx 2.31$ & $\approx 2.45$ 
    & $\approx 0.786$ & $\approx 0.447$     \\
    \hline  
    2 & 3/2 & $\approx 2.98$ & $\approx 2.72$ & $\approx 3.01$ 
    & $\approx 0.758$ & $\approx 0.448$     \\
    \hline  
\end{tabular}
\end{table}

\section{Doppler boost coefficient} 
\label{sec:boost}

Once we determined the effective radius of the emitting zone, it is natural to define the Doppler boost coefficient as 
\begin{equation}
    C_{_\Gamma} \equiv \frac{\delta \left( r=r_\mathrm{em} \right)}{\Gamma_\mathrm{sh}}    \ ,
\end{equation}
where the Doppler factor is calculated from Eq.~(\ref{delta_term}) at the distance equal to $r_\mathrm{em}$. For an adiabatic shock, when there exists analytic expression for $r_\mathrm{em}$ (see Eq.~\ref{R_em-adiabatic}), the above definition translates into 
\begin{equation} \label{C_Gamma}
    C_{_\Gamma}  = \frac{1}{2} \left(  2\sqrt{2} (m+1) - 2m -1 \right)^{\frac{m}{2(m+1)}}    \ .
\end{equation}
Otherwise, the value of $C_{_\Gamma}$ is to be evaluated following numerical determination of $r_\mathrm{em}$. 
Values of the Doppler boost coefficients $C_{_\Gamma}$ calculated for some special cases are given in Table~(\ref{tab:coefficients}). 
Note that \cite{SEDmodel}, who otherwise employed the coefficient values derived in this paper, used a two times smaller value for $C_{_\Gamma}$.
The updated analysis is not expected to reveal a qualitative difference.

There exists another possible definition of the Doppler boost coefficient, that involves averaging over entire equal arrival time surface. First, we calculate the total photon production rate from the bolometric luminosity, given by Eq.~(\ref{luminosity(r)_simple}):
\begin{equation} \label{Ndot_via_luminosity}
    \dot{N}_\mathrm{ph} (t_\mathrm{obs}) = \frac{L (t_\mathrm{obs})}{\langle h \nu_\mathrm{obs} \rangle}
    \simeq \frac{4\pi \epsilon_\mathrm{r} \rho_\mathrm{sh} c^3 R_\mathrm{sh}^2 \Gamma_\mathrm{sh}^4}{\langle h \nu_\mathrm{obs} \rangle}
   \,I_1\!\!\left( m,k \right)     \ ,
\end{equation}
where $\langle h \nu_\mathrm{obs} \rangle$ is the average energy of observed photons. Second, we calculate $\dot{N}_\mathrm{ph} (t_\mathrm{obs})$ directly:
\begin{equation} \label{full_Ndot}
    \dot{N}_\mathrm{ph} (t_\mathrm{obs}) = 2\pi \int_{-1}^{1} \frac{\lambda}{\langle h \nu \rangle} \delta^2 \diff{\cos\theta}     \, ,
\end{equation}
where $\langle h \nu \rangle$ is the average energy of radiated photons in the emitting zone's comoving frame. This expression is similar to Eq.~(\ref{full_luminosity}), differing from it by absence of one power of the Doppler factor, that took into account photons' blueshift. 
Following the same recipe as in 
Sect.~(\ref{sec:luminosity}), the above expression evaluates to
\begin{equation} \label{Ndot(r)_simple}
    \dot{N}_\mathrm{ph} (t_\mathrm{obs}) 
    \simeq \frac{4\pi \epsilon_\mathrm{r} \rho_\mathrm{sh} c^3 R_\mathrm{sh}^2 \Gamma_\mathrm{sh}^3 }{\langle h \nu \rangle}
    \,I_2\!\left( m,k \right) \ ,
\end{equation}
where
\begin{equation} \label{Ndot(r)_integral}
    I_2\!\left( m,k \right) =  \int_0^1     2 \sqrt{2}   \left( m+1 \right)^2  r^{1-k-\frac{3m}{2}}  \left( 2m + 1 +  r^{-(m+1)} \right)^{-2}     \diff{r}    \, . 
\end{equation}

Comparing Eqs.~(\ref{Ndot_via_luminosity}) and (\ref{Ndot(r)_simple}), we arrive at the alternative definition of the Doppler boost coefficient:
\begin{equation} \label{C_Gamma_alternative}
    \widetilde{C}_{_\Gamma} \equiv \frac{1}{\Gamma_\mathrm{sh}}
    \frac{\langle h \nu_\mathrm{obs}  \rangle}{\langle h \nu \rangle}     
    =     \frac{I_1\!\!\left( m,k \right)}{I_2\!\left( m,k \right)} \ .
\end{equation}
Usually either $I_1$, or $I_2$, or both do not have analytic representation. We calculate $\widetilde{C}_{_\Gamma}$ numerically for a few selected cases and compare them to the values of $C_{_\Gamma}$ in Table~(\ref{tab:C_Gamma_comparison}). There is a clear tendency for the values $\widetilde{C}_{_\Gamma}$ to be smaller than the values $C_{_\Gamma}$, but the difference is never large.

\begin{table} 
\caption{Comparison of local ($C_{_\Gamma}$) and brightness-weighted ($\widetilde{C}_{_\Gamma}$) values of Doppler boos coefficient.}
\label{tab:C_Gamma_comparison}
\begin{tabular}{cccc}
\hline
    $k$ & $m$ & $C_{_\Gamma}$ & $\widetilde{C}_{_\Gamma}$ \\
    \hline
    0 & 3 & $\approx 1.73$ & $\approx 1.52$     \\
    \hline  
    0 & 4 & $\approx 1.93$ & $\approx 1.64$     \\
    \hline  
    2 & 1 & $\approx 1.28$ & $\approx 1.07$     \\
    \hline  
    2 & 3/2 & $\approx 1.36$ & $\approx 1.12$     \\
    \hline  
\end{tabular}
\end{table}

\section{Time coefficient} 
\label{sec:time}

An observer moving with a fluid element next to the shock in the downstream observes the shock (i.e., upstream-facing border of the emitting zone) receding at the speed $c/3$. Define the position of downstream-facing border in such a way, that it also recedes from this fluid element at the speed $c/3$. 
Then, for an  observer in the lab frame, the radiation produced at the inner border (and being isotropic in the local frame) appears having two times smaller energy density compared to the radiation produced at the outer border. 
More distant fluid elements in the downstream recede faster, their radiation is more diluted due to smaller Lorentz boost and is more delayed. 
Note that if emission from the shock front reaches observer at time $t_{\rm obs}$ after the explosion, then emission from the downstream-facing border (defined as above) reaches the observer at time $2 t_{\rm obs}$.

A fluid element at the inner, downstream-facing, border of emitting zone has (by definition) the Lorentz factor  $\Gamma_\mathrm{in} = \Gamma_\mathrm{d}/\sqrt{2}=\Gamma_\mathrm{fr}/2$. Adiabatic blast wave solution of \cite{BlandfordMcKee} -- through Eqs.~66 and 67 in their paper -- relates the Lorentz factor of a fluid element to its coordinate (in terms of similarity variable $\chi$),
\begin{equation} \label{Gamma-chi}
    \Gamma_\mathrm{el} = \frac{\Gamma_\mathrm{fr}}{\sqrt{2 \chi_\mathrm{el}}} \, ,
\end{equation}
and provides expression for the mass enclosed inside the surface with coordinate $\chi_\mathrm{el}$ 
\begin{equation} \label{R-chi}
    M(\chi_\mathrm{el}) = \chi_\mathrm{el}^{\frac{k-3}{4-k}} M 
    \qquad \Rightarrow \qquad
    \chi_\mathrm{el} = \left( \frac{R}{R_1} \right)^{m+1}    \, ,
\end{equation}
where $k=3-m$ (adiabatic shock), $M \equiv M(\chi=1)$ is the total mass swept up by the shock, and $R_1$ position of the element at the moment when the shock traverses it. 

It follows from Eq.~(\ref{Gamma-chi}) that the coordinate of inner boundary of the emitting region is $\chi_\mathrm{in} = 2$. Consider a fluid element that enters the emitting region when $R=R_1$. It leaves the emitting region when $R=R_2=\chi_\mathrm{in}^{1/(m+1)} R_1$. 
While the shock propagates from $R_1$ to $R_2$, the fluid element crosses the entire emitting zone, from outer to inner boundary, and its Lorentz factor evolves as 
\begin{equation}
    \Gamma_\mathrm{el} 
    = \frac{\Gamma_1}{\sqrt{2}} \left( \frac{R}{R_1} \right)^{-m-1/2} \, .
\end{equation}
Here $\Gamma_1 \equiv \Gamma_\mathrm{fr} \left( R_1 \right)$; we used Eqs.~\ref{Gamma-chi}, \ref{R-chi} and the relation $\Gamma_\mathrm{fr} = \Gamma_1 \left( R/R_1 \right)^{-m/2}$.
So, the comoving-frame time that the fluid element spends inside the emitting zone is 
\begin{equation} \label{t_eff-general}
    t_\mathrm{eff} 
    = \frac{1}{c} \int_{R_1}^{R_2} \frac{\mathrm{d} R}{\Gamma_\mathrm{el}} 
    = \frac{\sqrt{2}}{m+3/2} \, \frac{R_1}{\Gamma_1 \, c} \left[ \left( \frac{R_2}{R_1} \right)^{m+3/2} -1 \right] \, .
\end{equation}

We choose such values of $R_1$ and $R_2$ that $\sqrt{R_1 R_2} = R_\mathrm{em}$. Together with the ratio $R_2/R_1 = 2^{1/(m+1)}$, this means that $R_1 = 2^{-1/\left[2(m+1)\right]} R_\mathrm{em}$. Substituting into Eq.~(\ref{t_eff-general}) the relation $R_1/\Gamma_1 = \left( R_1/R_\mathrm{sh} \right)^{1+m/2} R_\mathrm{sh}/\Gamma_\mathrm{sh}$ and then expressing $R_1$ via $R_\mathrm{em}$, we obtain
\begin{equation}
    t_\mathrm{eff} 
    = \frac{2}{2m+3} \, \frac{R_\mathrm{sh}}{\Gamma_\mathrm{sh} c}   
    \left( \frac{R_\mathrm{em}}{R_\mathrm{sh}} \right)^{\frac{m+2}{2}}
    2^{\frac{m+2}{4(m+1)}} 
    \left( 2 - 2^{-\frac{1}{2(m+1)}} \right) \, .
\end{equation}
Finally, using Eq.~(\ref{Rsh}) and the definition Eq.~(\ref{effective_time}), this gives 
\begin{equation} \label{C_t}
    C_\mathrm{t} 
    = \frac{4 \left( m+1 \right)}{2m+3}     
    \left( \frac{R_\mathrm{em}}{R_\mathrm{sh}} \right)^{\frac{m+2}{2}}
    2^{\frac{m+2}{4(m+1)}} 
    \left( 2 - 2^{-\frac{1}{2(m+1)}} \right) \, .
\end{equation}
Strictly speaking, the above expression is derived for adiabatic shocks, but we will use it also for partially radiative shocks taking the correct value of $R_\mathrm{em}/R_\mathrm{sh}$ ratio. Doing so we ignore modification of the downstream flow due to radiative losses, that would exceed the accuracy of the model (see also discussion in Sect.~\ref{sec:discuss}). The values of the time coefficient from Eq.~(\ref{C_t}) are listed in Table~(\ref{tab:coefficients}).

\section{Results and discussion} 
\label{sec:discuss}

In this paper we established correspondence between one-zone emission models, that take a single set of parameters, and the actual geometry of expanding relativistic shocks, where the parameters change along the surface of equal arrival time. Averaging over this surface results in a set of numeric coefficients entering the obvious scaling laws (see Eq.~\ref{CoefficientDefinitions}). The values that we find for these coefficients are summarized in Table~(\ref{tab:coefficients}) along with the values taken from few seminal papers. It is important to note large, almost reaching an order of magnitude, scatter in estimates of the coefficients by different authors. Therefore, using silently different sets of coefficients for numerical spectral models is very likely to lead to irreproducible results. For this reason, we urge the community to specify explicitly what coefficients are being used in spectral models. 

\begin{table*} 
\caption{Coefficients used in different one-zone afterglow  models. Note that in many cases these coefficients are introduced implicitly rather than explicitly in the corresponding papers. If a particular coefficient does not appear in the paper, neither explicitly nor implicitly, then it is not listed in the table.
The values of $m$ and $\epsilon_\mathrm{r}$ are related via Eq.~80 in \protect\cite{PartiallyRadiativeShock}.
}
\label{tab:coefficients}
\begin{tabular}{cccccccccc}
\hline
    Reference & Density profile & $m$ & $\epsilon_\mathrm{r}$ & $C_\mathrm{_E}^*$ 
    & $C_\mathrm{_L}$ & $C_\mathrm{_R}$ & $C_\mathrm{t}$ & $C_{_\Gamma}$ & $C_\mathrm{w}$\\
    \hline
    \cite{Afteglow2} & ISM ($k=0$) & 3 & $o(1)$ & 6/17 
    & $17/12\,^\dagger$ & $2$ & $1/\sqrt{2}$ & $1/\sqrt{2}$     \\
    \hline  
    \cite{Waxman97} & ISM ($k=0$) & 3 & $o(1)$ & 6/17  
    &  & $4\sqrt{2}$ &  & $2^{1/4}$ & $\approx 3$     \\
    \hline  
    \multirow{2}{*}{\cite{PanaitescuMeszaros98}}& wind ($k=2$) & 1 & $o(1)$ & 2/9  
    & & $\approx 3.1^\ddagger$ & & & $\approx 1.75$\\
    & ISM ($k=0$) & 3 & $o(1)$ & 6/17  
    & & $\approx 6.6^\ddagger$ & & & $\approx 2.93$\\
    \hline  
    \multirow{2}{*}{\cite{DaiLu98}}& wind ($k=2$) & 1 & $o(1)$ & 2/9  
    & & $4$ & $8\sqrt{2}/3$ &\\
    & ISM ($k=0$) & 3 & $o(1)$ & 6/17  
    & & $8$ & $16\sqrt{2}/5$ & \\
    \hline  
    \multirow{4}{*}{this work} & \multirow{2}{*}{wind ($k=2$)} & 1 & $o(1)$ & 2/9 
    & $9/8$ & $\approx 2.45$ & $\approx 1.16$ & $\approx 1.28$ & $\approx 1.75$ \\
    & & 3/2 & $\approx 0.155$ & $\approx 0.207$
    & $\approx 1.31$ &  $\approx 2.98$ & $\approx 0.97$ & $\approx 1.36$ & $\approx 2.08$  \\
    \cline{2-10}  
    &  \multirow{2}{*}{ISM ($k=0$)} & 3 & $o(1)$ & 6/17 
    & $17/16$ & $\approx 5.55$ & $\approx 0.96$ & $\approx 1.73$  & $\approx 2.93$ \\
    & & 4 & $\approx 0.165$ & $\approx 0.330$ 
    & $\approx 1.45$ & $\approx 6.87$ & $\approx 0.77$ & $\approx 1.93$ & $\approx 3.41$ \\
    \hline
\end{tabular}
\\
$^*$ We do not calculate this coefficient and instead take it from \cite{BlandfordMcKee} (for adiabatic case) or from \cite{PartiallyRadiativeShock} (for partially radiative case). \\
$^\dagger$ This coefficient does not appear in the paper, but in the slow-cooling case it can be calculated following the approach, that the authors used to normalize the distribution of emitting electrons. \\
$^\ddagger$ The paper presents different coefficients for slow-cooling and fast-cooling regimes, we take the numbers for their fast-cooling case.
\\
\end{table*}

We have demonstrated that calculating the set of effective coefficients in a self-consistent way results in well constrained estimates for the effective coefficients. 
Even in the cases where one can admit alternative yet reasonable definitions for the coefficients (see Sects.~\ref{sec:size} and \ref{sec:boost}), the numbers that correspond to different definitions diverge insignificantly. 
This reaffirms validity of our approach, where the effects arising from geometry and evolution of the expanding relativistic shock are taken into account through averaging the parameters of one-zone emission model over the surface of equal arrival time. 

Our analysis covers both commonly used cases of relativistic shocks expanding with negligible radiative losses into wind-like and constant density profiles, and further generalizes them for the more practical case of partially radiative shocks with constant fraction of radiative losses. 
Table~(\ref{tab:coefficients}) has four sets of coefficients, two for a shock expanding into wind-like density profile and two for a shock expanding into constant-density medium. One set from every pair is for adiabatic shock and another is for a selected value of the fraction of radiative losses, $\epsilon_\mathrm{r} \sim 0.15$, that we find realistic. The dependence of all coefficients on $\epsilon_\mathrm{r}$ is close to linear and there is no need in explicitly presenting coefficients for other values of $\epsilon_\mathrm{r}$, either in the table or in plots.

We finish with some notes about the assumptions that were used in this paper. 
The assumption that all the emission is produced close to the shock front holds for sufficiently energetic electrons, those that radiate their energy over timescale much smaller than $t_\mathrm{eff}$. 
The slow-cooling electrons radiate a small part of their energy and move with the flow far into downstream, gradually losing their energy for the flow expansion. 
To treat them properly, one needs to integrate over the downstream taking into account inhomogeneities in the flow parameters (such as magnetization) and in the electron distribution. 
Going into such details is hardly justified given poorly understood microphysics of relativistic shocks. Integration ignoring the inhomogeneities, on the other hand,
has little advantage over even simpler assumption that all the emission comes from a single location: it introduces comparable error while enforcing significantly more complex formalism. We therefore stick to this simplest yet reasonable assumption, and in application to slow-cooling electrons the parameter $t_\mathrm{eff}$ characterizes the fraction of energy they radiate rather than the extent of their distribution into downstream.

Our analysis holds for small radiative efficiency. Large $\epsilon_\mathrm{r} \sim 1$ would change jump conditions at the shock and the flow structure in the downstream, thus affecting the coefficients $C_\alpha$. 
Here we note that no theory predicts large radiative efficiency for relativistic shocks. Moreover, there is possible observational evidence for the opposite. 
Gamma-ray burst afterglows commonly exhibit power-law flux decay with time, i.e. $L \propto t_\mathrm{obs}^{-\alpha}$.
In almost every case $\alpha > 1$, and this is faster than $t_\mathrm{obs}^{-1}$ law expected from Eq.~(\ref{luminosity}) with $\epsilon_\mathrm{r}$ and $E_\mathrm{kin}$ unchanged. 
If this faster flux decrease is attributed to decreasing shock's energy, while $\epsilon_\mathrm{r}$ is constant, then $\alpha = 1 + (m+k-3)/(m+1)$ and for a typical $\alpha \approx 1.25$ the shock's deceleration law has exponent $m \approx 1.5$ (for wind-like external density profile) or $m \approx 4$ (for constant-density external medium). 
In both cases this translates into $\epsilon_\mathrm{r} \sim 0.15 \ll  1$ (see Eq.~80 in \cite{PartiallyRadiativeShock}).

Constancy of radiative efficiency $\epsilon_\mathrm{r}$ is not critical for our analysis. Generalization for $\epsilon_\mathrm{r} = \epsilon_\mathrm{r}(r)$ is straightforward --- one needs to keep $\epsilon_\mathrm{r}$ within the integrands rather than take it out as a constant multiplier. 
However, specific predictions for the dependence  $\epsilon_\mathrm{r}(r)$ are absent, and the assumption $\epsilon_\mathrm{r} = const$ is natural choice not only owing to its simplicity but also because it finds support in predictions of the pair balance model \citep{PairBalance}.


\section*{Acknowledgements}
This work was supported by the Russian Science Foundation under grant no. 21-12-00416.


\section*{Data Availability}

The data underlying this article are available in the article.



\bibliographystyle{mnras}
\bibliography{references} 

\label{lastpage}

\end{document}